\documentclass[aps,prb,twocolumn,groupedaddress]{revtex4}

\usepackage{graphicx}
\begin{document}

% Use the \preprint command to place your local institutional report
% number in the upper right-hand corner of the title page in preprint mode.
% Multiple \preprint commands are allowed.
% Use the 'preprintnumbers' class option to override journal defaults
% to display numbers if necessary
%\preprint{}

%\title{Transport properties  
%in disordered magnetic fields with a fixed sign}
\title{Quantum transport properties of two-dimensional systems 
in disordered magnetic fields with a fixed sign}

\author{Tohru Kawarabayashi}
%\email[]{Your e-mail address}
%\homepage[]{Your web page}
%\thanks{}
%\altaffiliation{}
\affiliation{Department of Physics, Toho University,
Miyama 2-2-1, Funabashi 274-8510, Japan}

\author{Tomi Ohtsuki}
%\email[]{Your e-mail address}
%\homepage[]{Your web page}
%\thanks{}
%\altaffiliation{}
\affiliation{Department of Physics, Sophia University,
Kioi-cho 7-1, Chiyoda-ku, Tokyo 102-8554, Japan}

\date{\today}

\begin{abstract}
Quantum transport in disordered magnetic fields 
is investigated numerically in two-dimensional systems.
In particular, the case where the mean and the fluctuation of 
disordered magnetic fields are of the same order is considered.
It is found that in the limit of weak disorder 
the conductivity exhibits a qualitatively different 
behavior from that in the conventional random magnetic fields with zero mean.
The conductivity is estimated by the equation of motion method and 
by the two-terminal Landauer formula.
It is demonstrated that the conductance 
stays on the order of 
$e^2/h$ even in the weak disorder limit.  
The present behavior can be 
interpreted in terms of the Drude formula.
The Shubnikov-de Haas oscillation is also observed in the weak disorder 
regime. 
\end{abstract}

% insert suggested PACS numbers in braces on next line
% insert suggested keywords - APS authors don't need to do this
%\keywords{}

%\maketitle must follow title, authors, abstract, \pacs, and \keywords
\maketitle

\section{Introduction}
Quantum transport in disordered magnetic fields in two dimensions(2D) 
has been studied widely by numerical and by analytical methods. 
A two-dimensional disordered system in random magnetic fields with zero mean
arises in a mean field theory of the fractional quantum Hall 
effect at filling factor $\nu=1/2$.\cite{HLR} The magneto transport around
$\nu = 1/2$ has then been often analyzed by models with random 
magnetic fields.\cite{HLR,KZ,KWAZ,AAMW,MPW,EMPW,MWEPW} 
The possibility of the 
Anderson transition in two-dimensional system in random magnetic fields
has also been studied extensively in the last 
decade.\cite{PZ,SN,AHK,AMW,YG,Verges,Furusaki} 
Although much work has been done to clarify 
whether or not the metallic states exist in such 
a system which belongs to the unitary universality class, 
the conclusions still remain controversial.
Systems in random magnetic fields show singularities at the 
band center in the conductance
fluctuations\cite{OSO} and in the density of states.\cite{Furusaki,MBF} 
These singularities 
are expected to be governed by the chiral symmetry of the 
random magnetic field system.
In recent experiments, two-dimensional electron systems in
random magnetic fields have been constructed and the magneto transport 
in such systems has been measured.\cite{Ando} 
The random magnetic fields with zero 
mean are realized by attaching small 
magnets on the layer parallel to the 2D electron gas in a semiconductor
heterostructure. It is then found that the magneto resistance 
exhibits  similar structure to that observed in the fractional
quantum Hall effect around $\nu = 1/2$.
It is thus important to understand the transport in random magnetic fields 
both theoretically and experimentally.

In the present paper, we focus on another aspect of the 
transport in disordered magnetic fields. 
In the previous paper,\cite{KOrg2d} 
we have investigated the magneto transport in  
random magnetic fields and found that 
the conductivity is 
insensitive to the magnitude of the fluctuation of the 
random magnetic fields if the mean value  
$\bar{B}$ of random magnetic fields 
is set to be of the same order of its fluctuation $\delta B$.
This is in contrast to the case of $\bar{B}=0$ where the conductivity 
diverges as $\propto (\delta B)^{-2}$ in the limit of 
$\delta B \rightarrow 0$.
It is thus an interesting problem to clarify whether the system exhibits 
this insensitivity even in the limit of zero magnetic fields.
We thus focus on the transport properties in the case where $\delta B \approx 
\bar{B}$, particularly in the limit of weak magnetic fields. 
The characteristic feature of such systems is that the magnetic fields
have almost the same sign. 
We therefore consider, as a simplified model, the two-dimensional 
system in  random magnetic fields
with a fixed sign in order to analyze the case $\delta B \approx \bar{B}$. 

Generally, it is expected that the 
scattering mechanism in random magnetic fields with a fixed sign is 
qualitatively different from that in the conventional 
random magnetic field with zero mean.
The system has, for example, no snake state near the zero magnetic field 
lines, which plays an important role in the semi-classical theory of the 
transport in the conventional random magnetic fields.\cite{EMPW,MWEPW}
It would thus  be useful to study the present system also for the further  
understanding of the quantum transport in 
the conventional random magnetic fields. 
On the other hand, 
in the strong magnetic field limit $\bar{B} \gg \delta B$ 
it has been argued and demonstrated that the statistical properties belong 
to the same universality class as the quantum Hall 
transition in two dimensions.\cite{Huckestein,Yakubo3}
The float of extended states toward the limit of the weak magnetic field
$\bar{B} \ll \delta B$ has then become an important issue.\cite{Yakubo3,CYH} 
Clarifying the transport properties of 
the present case ($\bar{B} \approx \delta B$), particularly
in the limit of weak magnetic fields, would be important 
also in considering the connection between these two limiting cases.

We evaluate numerically the 
conductance in random magnetic fields with a fixed sign by  
the equation of motion method and by the Landauer formula. 
In the equation of motion method, the conductance 
is obtained by examining the electron diffusion directly.
This method has an advantage that it applies to very large 
systems. On the other hand, in the case of the Landauer formula,
large numbers of samples can be considered, although 
system-sizes are limited. 
With these two independent numerical method,
we calculate the longitudinal conductivity and 
find that the conductance exhibits a qualitatively different
behavior from that in the conventional random magnetic fields,
particularly, in the limit of weak disorder. 
We also show that the present observations are not changed significantly  
by introducing a small number of magnetic fields with the 
opposite sign.

\section{Model}
We adopt a model described by the following Hamiltonian on the 
square lattice  
\begin{equation}
 H = \sum_{<i,j>} V \exp({\rm i} \theta_{i,j}) C_i^{\dagger}
 C_j ,
\end{equation}
where $C_i^{\dagger}(C_i)$ denotes the creation(annihilation) operator 
of an electron on the site $i$.  
The phases $\{ \theta_{i,j} \}$ are related to the disordered 
magnetic fluxes 
$\{ \phi_i \}$ through the plaquette 
$(i,i+\hat{x},i+\hat{x}+\hat{y},i+\hat{y})$ as 
\begin{equation}
 \theta_{i,i+\hat{x}} + \theta_{i+\hat{x},i+\hat{x}+\hat{y}}+
 \theta_{i+\hat{x}+\hat{y},i+\hat{y}} + \theta_{i+\hat{y},i}
 = -2\pi \phi_i / \phi_0
\end{equation}
where $\phi_0 = h/|e|$ stands for the unit flux and $\hat{x}(\hat{y})$ 
denotes the unit vector in the $x(y)$-direction.
The fluxes $\{ \phi_i \}$ are assumed to be distributed independently 
in each plaquette. The probability distribution $P(\phi)$ of the flux $\phi$ 
is given by
\begin{equation}
 P( \phi ) = \left\{ \begin{array}{ll}
                     1/h_{\rm rf} & {\rm for }\quad 0 \leq \phi /\phi_0 
                     \leq h_{\rm rf} \\
                      0 & {\rm otherwise}
                     \end{array}  \right.  .
\end{equation}
The mean and the variance of the distribution are accordingly given by
\begin{equation}
 \langle \phi_i \rangle = \frac{h_{\rm rf}}{2}\phi_0 \quad {\rm and} \quad 
 \langle \phi_i \phi_j \rangle - 
 \langle \phi_i \rangle \langle \phi_j \rangle
 = \frac{h_{\rm rf}^2}{12}\phi_0^2 
 \delta_{i,j} , 
\end{equation}
respectively.

\section{ELECTRON DIFFUSION}

In order to verify that the above system has a 
diffusive regime, we first examine the diffusion 
of electrons by means of the equation of motion method.
For this, we numerically 
solve the time-dependent Schr\"{o}dinger equation by making use of 
the decomposition formula for 
exponential operators.\cite{Suzuki} 
The basic formula used in the present paper 
is the forth order formula
\begin{equation}
 \exp ( x[A_1 + \cdots + A_n]) = S(xp)^2 S(x(1-4p))S(xp)^2 + O(x^5),
\end{equation}
where 
\begin{equation}
 S(x) = {\rm e}^{xA_1/2}\cdots {\rm e}^{xA_{n-1}/2}{\rm e}^{xA_n}
 {\rm e}^{xA_{n-1}/2}\cdots {\rm e}^{xA_1/2}.
\end{equation}
The parameter $p$ is given by $p=(4-4^{1/3})^{-1}$ and $A_1, \ldots , A_n$
are arbitrary operators.
We divide the Hamiltonian into five parts as in the 
previous papers\cite{KO} so that each part is represented 
as the product of $2\times 2$ submatrices.
By applying this formula to the time evolution operator $U(t)\equiv 
\exp(-{\rm i}
H t/\hbar)$, we obtain 
\begin{equation}
 U(\delta t) = U_2(p \delta t)^2 U_2((1-4p)\delta t) U_2(p \delta t)^2 +
 O(\delta t^5)
\end{equation} 
with
\begin{equation}
 U_2(x) = {\rm e}^{xH_1/2}\cdots {\rm e}^{xH_4/2}{\rm e}^{xH_5}
 {\rm e}^{xH_4/2}\cdots {\rm e}^{xH_1/2} ,
\end{equation}
where $H=H_1 + \cdots + H_5$. It is to be noted that the $U_2$ can 
be expressed in an  analytical form while the original evolution operator 
$U$ can not be evaluated exactly without performing the 
exact diagonalization of the whole system. 
This method has already 
been applied to the various cases of 2D\cite{KOrg2d,KO,KO2} and 
3D\cite{OK} disordered systems. 

Numerical calculations are performed in the system of the 
size $999 \times 999$ with the fixed boundary condition. 
All the length scales are measured in
units of the lattice constant $a$.
To prepare the initial wave packet with energy $E$, we numerically 
diagonalize the subsystem ($21 \times 21$) 
located at the center of the whole system and 
take  as the initial 
wave function the eigenstate whose energy is closest to $E$. 
The single time step $\delta t$ 
is set to be $\delta t = 0.2 \hbar/V$
and at least five realizations of random magnetic fields are considered.
With this time step, the fluctuations of the expectation value of 
the Hamiltonian is safely neglected throughout the present 
simulation $(t \leq 2000 \hbar/V)$.
We observe the second moment defined by 
\begin{equation}
 \langle r^2 (t)\rangle_c \equiv \langle r^2 (t)\rangle -
 \langle r (t)\rangle^2
\end{equation}
with 
\begin{equation}
 \langle r^n(t) \rangle = 
 \int r^n {\rm d}\Omega r^{d-1}{\rm d}r |\psi(\mathbf{r},t)|^2
\end{equation}
where $\psi(\mathbf{r},t)$ denotes the wave function at time $t$.
In the diffusive regime, the second moment is expected to 
grow in proportion to $t$
\begin{equation}
 \langle r^2 \rangle_c = 2dDt ,
\end{equation}
where the diffusion coefficient is denoted by $D$ and $d$ is
the dimensionality of the system. The 
diffusion coefficient $D$ is related to the conductivity 
by the Einstein relation $\sigma = e^2 D \rho$. Here 
$\rho$ denotes the density of states. 
It is estimated by the Green function 
method\cite{SKM} for strips with the width up to $30$.  
 
The second moment $\langle r^2 \rangle_c$ for $h_{\rm rf}=0.04$ as a function 
of time $t$ is shown in Fig. 1. The fermi energy is assumed to be 
$E/V=0.5$. It is clearly 
seen that, above a certain length
scale, the second moment grows in proportion to the time $t$.
In this regime, it is natural to expect that the system is diffusive.  
The length scale, above which the diffusive behavior is observed, 
must be related to the typical cyclotron radius of the 
present system. Since the fermi energy is fixed close to the band center,
it would vary almost inversely proportional to  $h_{\rm rf}$. 
In fact, the length scales are estimated as $\sim 38$, $\sim 49$, and 
$\sim 70$ for $h_{\rm rf}=0.04$, $0.03$ and $0.02$, respectively. 
Below these length scales, the system is in a ballistic regime.
In estimating the diffusion coefficient, we discard the data in the 
ballistic regime. 
It should be noted that, in the present regime of $h_{\rm rf}$, 
the effective mean free path due to 
the fluctuation of magnetic fields is much larger than the cyclotron 
radius determined by the mean value of the magnetic fields.
For instance, for the case of $h_{\rm rf}=0.02$, 
the effective mean free path is estimated 
to be around $2100a$,\cite{KOrg2d} 
whereas the cyclotron radius is $40a$.   

\begin{figure}
\includegraphics[scale=0.6]{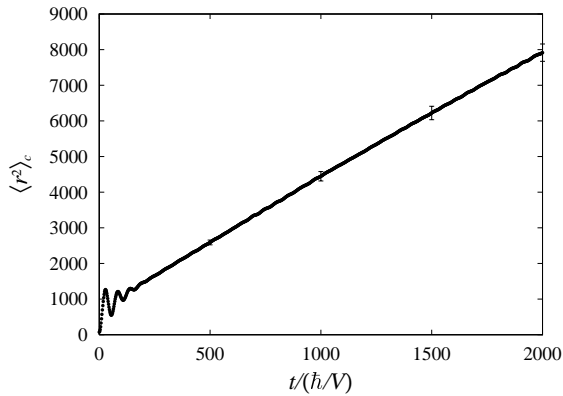}
\caption{The second moment averaged over 
10 realization of random magnetic 
fields as a function of time $t$ for 
$E/V = 0.5$ and $h_{\rm rf}=0.04$. The fluctuation around the 
mean value is plotted at $t/(\hbar /V)=$ $500$, $1000$,
$1500$ and $2000$. \label{fig1}}
\end{figure}

The estimated conductivity is plotted as 
a function of $h_{\rm rf}$ in Fig. 2.
It is rather remarkable that the conductivity stays 
on the order of the conductance quantum $e^2/h$ even for 
the small values of $h_{\rm rf}$. This feature is in contrast to 
the case of the conventional random magnetic fields, where 
the conductivity is likely to diverge inversely proportional to 
the square of the fluctuation of the random magnetic fields.
These facts mean that, provided that the sign of the magnetic fields is 
fixed, the conductivity is 
insensitive to the strength of its 
fluctuation, and implies  that the zero field limit $h_{\rm rf} \rightarrow 0$
does not coincide with the $h_{\rm rf}=0$ case and thus is a singular limit.

\begin{figure}
\includegraphics[scale=0.6]{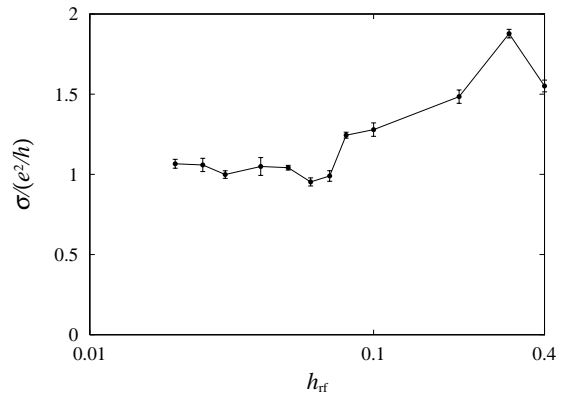}
\caption{The conductivity as a function of $h_{\rm rf}$ for 
$E/V = 0.5$. For $h_{\rm rf} < 0.1$, the conductivity stays 
on the order of the conductance quantum $e^2/h$. \label{fig2}}
\end{figure}

\section{CONDUCTANCE}

In order to investigate this small conductivity in the weak 
fluctuation regime, 
we examine also the conductance in the two-terminal geometry based on 
the Landauer formula 
\begin{equation}
 G = \frac{e^2}{h} {\rm Tr}( T T^{\dagger}),
\end{equation}
where $T$ denotes the transmission matrix through the system.
We consider the $L$ by $L$ square system 
with two leads connected to both sides of the system.
No magnetic field is assumed in the leads. 
For this system, we adopt the transfer matrix 
formalism \cite{Pendry} for evaluating the transmission coefficient 
numerically.
The maximum system-size considered in the present work is 160 by 160 and the 
average over 1000 realizations of random field configurations is performed.
Since the mean value of the magnetic fields is not zero in the present system,
the effect of the edge states is, in general, important for the 
transport properties. We therefore 
estimate the conductance both with the fixed boundary conditions and with
periodic boundary conditions in the transverse direction.

In Fig. 3, the conductances for the fixed boundary conditions for 
various system-sizes are shown as a function of $h_{\rm rf}$.
It is clearly seen that the conductance is likely to diverge as 
$h_{\rm rf}\rightarrow 0$, which is qualitatively different from the 
behavior obtained by the analysis of the electron diffusion.
It is also seen that the value of the conductivity 
is much larger than that estimated from the electron diffusion.
It is expected that these differences come from the fact that 
the edge states, which are absent in the simulation of 
the electron diffusion, exist in the case of the fixed boundary conditions. 
The observation that the conductance is almost inversely proportional to
$h_{\rm rf}$ (Fig. 3) suggests that it  
is the manifestation of the fact that the number of 
edge states increases as $h_{\rm rf}^{-1}$.

\begin{figure}
\includegraphics[scale=0.6]{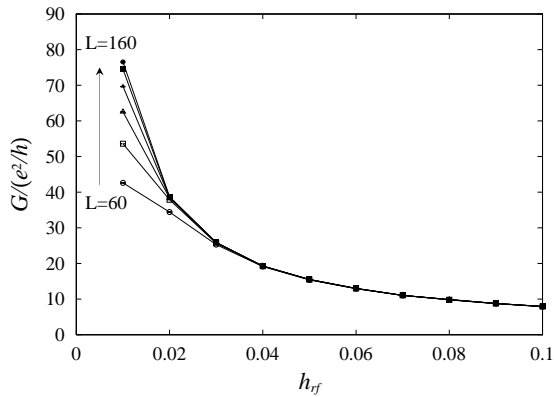}
\caption{The conductance under the fixed boundary conditions for 
$E/V = 0.5$ and $L=60$(open circles), $80$(open squares), 
$100$(open triangles), $120$(solid triangles), $140$(solid squares) and 
$160$(solid circles). The average over 1000 realizations of random fields is
performed. \label{fig3}}
\end{figure}

\begin{figure}
\includegraphics[scale=0.6]{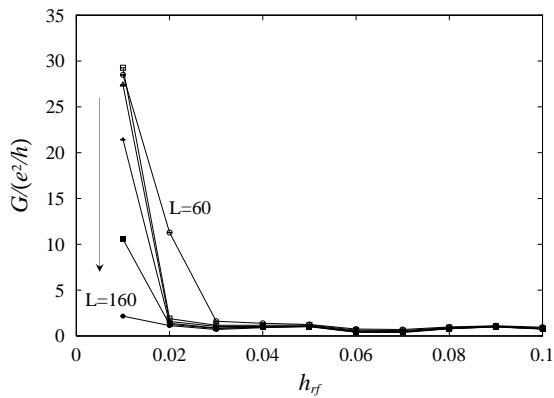}
\caption{The conductance under the periodic boundary conditions for 
$E/V = 0.5$ and $L=60$(open circles), $80$(open squares), 
$100$(open triangles), $120$(solid triangles), $140$(solid squares) and 
$160$(solid circles). The average over 1000 realizations of random fields is 
performed. \label{fig4}}
\end{figure}

In order to remove the effect of edge states, we evaluate the conductivity 
under the periodic boundary conditions in the transverse 
direction (Fig. 4).
It is seen that, in contrast to the 
case of the fixed boundary conditions, the conductance stays on the 
order of the conductance quantum even for the small values of 
$h_{\rm rf}$. It should be emphasized 
that for small $h_{\rm rf}$ 
the conductance decreases as the size of the system is increased.
We thus recover the results obtained from the 
electron diffusion. This suggests that it is essential to 
remove the effect of edge states for observing the small 
conductance
in the random magnetic fields with a fixed sign.

In the case of the periodic boundary conditions, the total flux 
through the system is adjusted to be an integer multiple of the 
flux quantum $h/|e|$ in order to ensure that the flux through the leads to 
be zero. Practically, we divide the excess flux into small 
pieces and subtract these pieces randomly from the fluxes already 
distributed in the system
as long as they are positive, so that the above condition is 
satisfied. This procedure produces 
a weak correlation in random fluxes which should be negligible for large 
system sizes.

\begin{figure}
\includegraphics[scale=0.6]{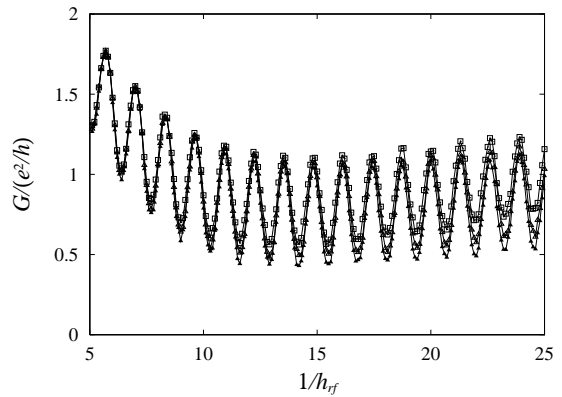}
\caption{The conductance under the periodic boundary conditions for 
$E/V = 0.5$ and $L=80$(open squares), 
$100$(open triangles), $120$(solid triangles) as a function 
of $1/h_{\rm rf}$. The period in $1/h_{\rm rf}$ is estimated to be 
$1.33$. \label{fig5}}
\end{figure}

More detailed calculations of the conductance for the 
periodic boundary conditions  are shown in Fig. 5.
It is clearly seen that the conductance oscillates periodically 
as a function of $1/h_{\rm rf}$ and its period is independent of the 
system-sizes. 
From our numerical data (Fig. 5), the period in $1/h_{\rm rf}$ 
is estimated to be $\Delta(1/h_{\rm rf})=1.33$. 
The same type of oscillation is 
observed also for $E/V=0.3$ and for $E/V = 0.7$ with different periods. 
The periods in $1/h_{\rm rf}$
are estimated to be $1.18$ and $1.41$, respectively.
Here we show that
these oscillations can be understood as the Shubnikov-de Haas
effect.\cite{Kittel}  
In general, the period
of the Shubnikov-de Haas effect for tight-binding lattice 
models is given by
\begin{equation}
 \Delta (1/\tilde{\phi}_{\rm ext} ) = \frac{4\pi^2}{a^2 S}, \quad 
 \tilde{\phi}_{\rm ext} \equiv \phi_{\rm ext}/\phi_0 
 \label{SdH}
\end{equation}
where $\phi_{\rm ext}$ denotes the external flux through the plaquette and 
$S$ the area of the fermi surface of the two-dimensional 
electron system. By evaluating $S$ in the system without magnetic 
fields, we obtain $\Delta (1/\tilde{\phi}_{\rm ext} ) = 2.36$, 
$2.59$ and $2.83$
for $E/V=0.3$, $0.5$ and $0.7$, respectively.
Since, in the present system, the mean value of the magnetic field 
$\langle \phi /\phi_0 \rangle $ is given by $h_{\rm rf}/2$, 
we can identify
$\tilde{\phi}_{\rm ext}$ to be $h_{\rm rf}/2$. We then find the 
relation that 
$\Delta (1/\tilde{\phi}_{\rm ext} ) = 2 \Delta (1/h_{\rm rf})$.
With this relation, it is clear that the periods of the  
Shubnikov-de Haas effect evaluated by eq.(\ref{SdH}) 
are in excellent agreement with those of the oscillation 
observed in our numerical data. It is rather remarkable that 
the present system exhibits such a clear Shubnikov-de Haas effect
even though the flux distributes uniformly from $0$ to $h_{\rm rf}$. 
A smaller oscillation than that of the conductance  
is observed in the density of states (Fig. 6), which 
seems to be consistent with the analytical results \cite{EMPW,TAndo} for
the conventional Shubnikov-de Haas oscillation of  
2D systems in magnetic fields.

\begin{figure}
\includegraphics[scale=0.6]{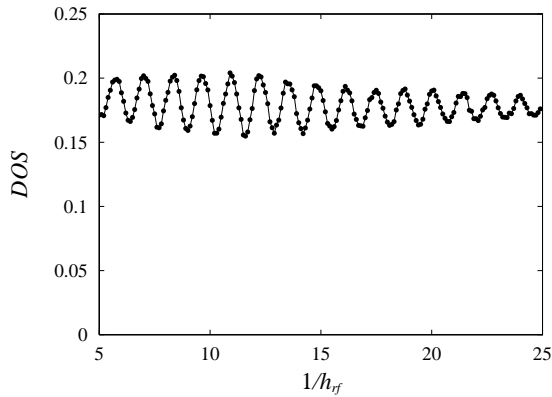}
\caption{The density of states for 
$E/V = 0.5$.\label{fig6}}
\end{figure}

Apart from this oscillation, the conductance shows 
a smooth cross over from the small $h_{\rm rf}$ regime to the 
$h_{\rm rf} = 1$ limit. It is to be noted that due to the 
periodicity of the phases  $\theta_{i,j}$ in the hamitonian, 
the $h_{\rm rf}=1$ limit is 
identical to the case of the conventional random magnetic field with 
zero mean. 
Our numerical data suggest that this cross over takes place around 
$h_{\rm rf}=0.1 \sim 0.2$. With this cross over,
the conductance is increased to a value $\sim 1.5 e^2/h$ (Fig. 5) as 
$h_{\rm rf}$ is increased.
This enhancement of conductance in the regime $h_{\rm rf} > 0.1$ is 
also observed in the the conductivity estimated by the 
equation of motion method (Fig. 2).

\section{Discussion}
We have investigated the longitudinal conductivity in the two-dimensional 
disordered magnetic fields with a fixed sign.
It has been demonstrated that the system shows a slow diffusion and
has accordingly a small conductance which is  
on the order of $e^2/h$ and is insensitive to 
the magnitude $h_{\rm rf}$ of the random fields.
It would be useful to interpret these results in terms of 
the Drude formula $\sigma = \sigma_0 /(1+\omega_c^2 \tau^2)$, 
where $\omega_c$ and $\tau$ denote the 
cyclotron frequency and the relaxation time, respectively.
The conductivity for $\omega_c = 0$ is denoted by $\sigma_0$. 
Here it should be kept in mind that the present system is equivalent to 
the system in the conventional random magnetic field distributed 
from $-h_{\rm rf}/2$ to $h_{\rm rf}/2$, plus an additional uniform 
magnetic field $h_{\rm rf}/2$. Namely, we can regard it as the conventional 
random field system in an external uniform magnetic field.
From this point of view, we assume that the relaxation time $\tau$ is 
determined simply by the scattering due to the 
conventional random fields with 
zero mean and is insensitive to the 
additional uniform magnetic field. We also assume that the 
effect of the additional uniform magnetic field appears only in 
$\omega_c$. It is then deduced \cite{EMPW, MWEPW} 
that $\sigma_0 \propto \tau \propto 
h_{\rm rf}^{-2}$ and $\omega_c \propto h_{\rm rf}$, and therefore that
for $h_{\rm rf} \ll 1$,   
\begin{equation}
 \sigma /(e^2/h)= \frac{\sigma_0/(e^2/h)}{(1+\omega_c^2 \tau^2)} \approx 
 \frac{A_1 h_{\rm rf}^{-2}}{(1+A_2 h_{\rm rf}^{-2})},
\end{equation} 
where $A_1$ and $A_2$ are constants independent of $h_{\rm rf}$.
If we take the limit as $h_{\rm rf} \rightarrow 0$, we obtain 
that $\sigma/(e^2/h) \rightarrow A_1/A_2$. 
The parameters $A_1$ and $A_2$
are basically determined by the fermi energy $E$.\cite{EMPW,MWEPW} 
For instance, 
$A_1$ has been estimated to be $0.95$ for $E/V=0.5$.\cite{KOrg2d}
The present argument based on the Drude formula, though it 
does not account for the Shubnikov-de Haas effect, 
seems to account for the fact that for $h_{\rm rf} \ll 1$ 
the conductance stays on the order of $e^2/h$ and is 
insensitive to $h_{\rm rf}$. In this context, the feature specific to the 
present system is that both $\tau$ and 
$(\tau \omega_c)^2$ scales in the same way 
as $h_{\rm rf}^{-2}$.

Finally, we emphasize that these present transport properties are 
not restricted to the case of the strictly positive 
random magnetic fields. These properties can be 
observed in the case where the random magnetic fields are almost 
positive. In Fig. 7, the conductance is shown for the case where 
the random fluxes $\{ \phi_i/\phi_0 \}$ are distributed in 
the range $[-0.05h_{\rm rf}, 0.95h_{\rm rf}]$. It is clear 
that the Shubnikov-de Haas effect is also observed even though the 
magnetic fields are not strictly positive. This suggests
that these properties are general features of systems in 
the random magnetic fields whose mean and fluctuation 
are of the same order.

\begin{figure}
\includegraphics[scale=0.6]{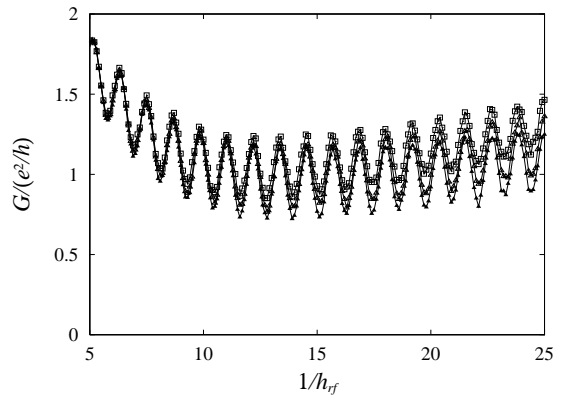}
\caption{The conductance for $E/V=0.5$ in the case where 
the random fluxes $\phi/\phi_0$ are distributed in the 
range $[ -0.05h_{\rm rf}, 0.95h_{\rm rf} ]$.
Open squares, open triangles, solid triangles represent 
$L=80$, $100$ and  $120$, respectively.\label{fig7}}
\end{figure}

In summary, we have studied numerically the transport properties 
in two-dimensional random magnetic fields with a fixed sign.
In particular, 
the conductivity in the limit of small fluctuation has been investigated.  
In the simulation of the diffusion of electron wave functions, it is 
observed that the system shows diffusive behavior in larger length-scales 
than the typical cyclotron radius. The diffusion turns out to be 
very slow and  
the conductivity evaluated from the diffusion coefficient stays on 
the order of the conductance quantum insensitive to the magnitude of 
fluctuation of the magnetic fields.    
This behavior is also observed in the two-terminal conductance 
evaluated by means of the Landauer formula
under the periodic boundary conditions.
A clear Shubnikov-de Haas effect has been observed for a wide 
range of $h_{\rm rf}$.
For a system with edge states, the contribution to the conductance 
from  the edge channels 
is much larger than that from the bulk, and hence 
these properties are not observable. In this sense, these 
peculiar features 
are the bulk properties of the positive random magnetic fields.
It has been argued that 
the singularity in the limit of $h_{\rm rf}\rightarrow 0$ can be 
understood within the framework of the Drude formula.

% If you have acknowledgments, this puts in the proper section head.
\begin{acknowledgments}
The authors thank Y.Ono, B.Kramer, and S.Kettemann for valuable 
discussions. Numerical calculations were partly performed by the facilities 
of the Supercomputer Center, Institute for Solid State Physics, 
University of Tokyo. 
\end{acknowledgments}

% Create the reference section using BibTeX:
%\bibliography{apssamp.bib}

\begin{thebibliography}{10}
%
\bibitem{HLR}
B. I. Halperin, P. A. Lee, and N. Read, Phys. Rev. B {\bf 47},  7312  (1993).
\bibitem{KZ}
V. Kalmeyer and S.C. Zhang, Phys. Rev. B {\bf 46}, 9889 (1992).
\bibitem{KWAZ} V. Kalmeyer, D. Wei, D.P. Arovas  and S. Zhang,
Phys. Rev. B {\bf 48}, 11095 (1993).
\bibitem{AAMW} A.G. Aronov, E. Altshuler, A.D. Mirlin and P. W\"{o}lfle,
Phys. Rev. B {\bf 52}, 4708 (1995).
\bibitem{MPW}
A.D. Mirlin, D.G. Polyakov and P. W\"{o}lfle, Phys. Rev. Lett. {\bf 80},
2429 (1998).
\bibitem{EMPW}
F. Evers, A. D. Mirlin, D. G. Polyakov, and P. W\"{o}lfle, Phys. Rev. 
B{\bf 60},  8951  (1999).
\bibitem{MWEPW}
A.D. Mirlin, J. Wilke, F. Evers, D.G. Polyakov and P. W\"{o}lfle, 
Phys. Rev. Lett. {\bf 83}, 2801 (1999).

\bibitem{PZ}
C. Pryor and A Zee, Phys. Rev. B {\bf 46}, 3116 (1992).
\bibitem{SN}
T. Sugiyama and N. Nagaosa, Phys. Rev. Lett. {\bf 70}, 1980 (1993).
\bibitem{AHK}
Y. Avishai, Y. Hatsugai and M. Kohmoto, Phys. Rev. B {\bf 47}, 9561 (1993).
\bibitem{AMW} A.G. Aronov, A.D. Mirlin and P. W\"{o}lfle, Phys.
Rev. B{\bf 49}, 16609 (1994).
\bibitem{YG}
K. Yakubo and Y. Goto, Phys. Rev. B {\bf 54}, 13432 (1996).
\bibitem{Verges}
J.A. Verg\'{e}s, Phys. Rev. B {\bf 57}, 870 (1998).
\bibitem{Furusaki} 
A. Furusaki, Phys. Rev. Lett. {\bf 82},  604  (1999).

\bibitem{OSO} T. Ohtsuki, K. Slevin and Y. Ono, J. Phys. Soc. Jpn. 
{\bf 62}, 3979 (1993).

\bibitem{MBF} 
C. Mudry and P. W. Brouwer, and A. Furusaki, Phys. Rev. B {\bf 59},  13221  
(1999).

\bibitem{Ando}
M. Ando, A. Endo, S. Katsumoto, and Y. Iye, Physica B{\bf 284}-{\bf 288},
1900 (2000); M. Ando, Thesis (University of Tokyo, 2001).


\bibitem{KOrg2d} T. Kawarabayashi and T. Ohtsuki, Phys. Rev. B{\bf 67}, 
165309 (2003).


\bibitem{Huckestein} B. Huckestein, Phys. Rev. B{\bf 53}, 3650 (1996).

\bibitem{Yakubo3} K. Yakubo, Phys. Rev. B{\bf 62}, 16756 (2000).

\bibitem{CYH} M.C. Chang, M.F. Yang and T.M. Hong, Phys. Rev. B{\bf 56},
3602 (1997).

\bibitem{Suzuki}
M. Suzuki, Phys. Lett. A {\bf 146}, 319 (1990). 

\bibitem{KO} T. Kawarabayashi and T. Ohtsuki, Phys. Rev. B{\bf 51},
10897 (1995).
\bibitem{KO2} T. Kawarabayashi and T. Ohtsuki, Phys. Rev. B{\bf 53},
6975 (1996).
\bibitem{OK} T. Ohtsuki and T. Kawarabayashi, J. Phys. Soc. Jpn. {\bf 66},
314 (1997).

\bibitem{SKM} L. Schweitzer, B. Kramer and A. MacKinnon, J. Phys. C{\bf 17},
4111 (1984).

\bibitem{Pendry} J. B. Pendry, A. MacKinnon and P. J. Roberts, 
Proc. R. Soc. Lond. A {\bf 437}, 67 (1992).

\bibitem{Kittel} C. Kittel, {\it Quantum Theory of Solids} (John Wiley \&
Sons, 1987).

\bibitem{TAndo} T. Ando, J. Phys. Soc. Jpn. {\bf 37}, 1233 (1974).



\end{thebibliography}

\end{document}